\documentstyle[preprint,aps]{revtex}
\begin{document}
\draft
\preprint{}
\title{Topological Electric Charge}
\author{D. Singleton}
\address{Department of Physics, University of Virginia, 
Charlottesville, VA 22901}
\date{\today}
\maketitle
\begin{abstract}
By treating magnetic charge as a gauge symmetry through the 
introduction of a ``magnetic'' pseudo four-vector potential, it is 
shown that it is possible, using the 't Hooft-Polyakov construction,
to obtain a topological electric charge. The mass of this electrically
charged particle is found to be on the order of $ {1 \over 137} M_W$
as opposed to the much larger mass (on the order of $137 M_W$) of the
magnetic soliton. Some model building possibilities are discussed 
\end{abstract}
\newpage
\narrowtext
\section{Introduction}
Using a non-Abelian gauge field coupled to a 
self-interacting scalar field, 't Hooft and 
Polyakov \cite{thooft} have shown that it is possible to construct a 
topological soliton which has a magnetic charge.
This magnetic monopole is guaranteed to be stable because of the
nontrivial topology of the vacuum expectation value (VEV) of the scalar
field. This strange vacuum configuration was called the ``hedgehog''
solution by Polyakov since the direction in isospin space in which
the VEV points is linked to the radial direction of ordinary space.
This magnetic soliton has several unique properties. First its
magnetic charge is not a Noether charge ({\it i.e.} it is not linked
with any apparent symmetry), but is a topological charge,
which owes its existence to the unusual vacuum of the
scalar field. Second the monopole has no singularites in its
fields (so long as the self-coupling, $\lambda$, of the scalar
field is not zero). Finally, and unfortunately,
it is estimated to have a mass on the order of $137 M_W$ (where
$M_W$ is the mass of the gauge bosons after symmetry breaking). If
$M_W$ is of the order of the electroweak gauge bosons ($\sim
100$ GeV) then seeing such magnetically charged objects is out of 
the question for any current or planned accelerator. The 
monopoles large mass comes about because of the value
of the gauge coupling of the electric $U(1)$ symmetry, 
which is embedded in the original non-Abelian
gauge theory. The gauge coupling of the original non-Abelian
gauge symmetry must be chosen to have the value of the electromagnetic
coupling $e$. This leads to the monopole having a large magnetic
charge of ${4 \pi\over e}$, and a large mass, on the order of 
${4 \pi \over e^2} M_W \approx 137 M_W$.

It might be asked if it is possible to construct such topologically 
stable solitons which have the far field of an electric charge.
Julia and Zee \cite{zee} have found field configurations 
which carry {\it both} magnetic and electric charge, which are
called dyons. However the electric charge can not exist without an
accompanying magnetic charge, and the stability arguments that apply
to the purely magnetic solution do not apply to the dyonic solution
(although there are plausiblity arguments for its stability). Purely
electrically charged solitons would be of 
more interest than magnetically charged solitons since 
electrically charged particles are much more common. The 
reason for thinking that electric solitons are possible is
the dual symmetry \cite{jackson} between electric
and magnetic quantities of Maxwell's equations
\begin{eqnarray}
\label{dual1} 
{\bf E} \rightarrow cos \theta {\bf E} + sin \theta {\bf B} \nonumber \\
{\bf B} \rightarrow -sin \theta {\bf E} + cos \theta {\bf B} 
\end{eqnarray}
and
\begin{eqnarray}
\label{dual2}
\rho _e \rightarrow  cos \theta \rho _e + 
sin \theta \rho _m  \; \; \; \; \;
{\bf J} _e \rightarrow cos \theta {\bf J} _e + 
sin \theta {\bf J} _m  \nonumber \\
\rho _m \rightarrow - sin \theta \rho _e + 
cos \theta \rho _m \; \; \; \; \;
{\bf J} _m \rightarrow -sin \theta {\bf J}_e + 
cos \theta {\bf J} _m 
\end{eqnarray}
where $\rho _{e(m)}$ and ${\bf J} _{e(m)}$ are the electric
(magnetic) charge and current densities. Thus given a particle
with a certain electric and magnetic charge it is possible 
to use this dual symmetry to ``rotate'' the two charges so that
the particle ends up with a different electric and
magnetic charge. By properly chosing the angle $\theta$ a 
particle can be made to carry only electric charge or only magnetic 
charge. The ability to altogether transform away one type of
charge holds only if all particles have the same ratio of
electric to magnetic charge, since the dual transformation
of Eq. (\ref{dual2}) is global.

Based on the dual symmetry between electric and magnetic quantities
it should be possible to find a topological electric soliton
by using a duality transformation on the 
the magnetic soliton. However the monopole solution is in 
terms of the gauge potentials
rather than the {\bf E} and {\bf B} fields, 
which are involved in the dual symmetry of Eq. ({\ref{dual1}). 
When {\bf E} and {\bf B} are written in terms of the
four-vector potential, $A_{\mu}$, ($E_i = \partial ^i A^0
- \partial ^0 A^i$ and $B_i = - \partial ^j A^k + \partial ^k
A^j$) it appears impossible to implement the dual symmetry
in terms of the potentials.

It has been shown by various authors \cite{cabbibo}, 
\cite{rohrlich}, \cite{zwang} that Maxwell's equations with
magnetic and electric charge can be dealt with by
introducing a second, pseudo four-vector potential, $C_{\mu}$,
in addition to the usual four-vector potential, $A_{\mu}$ 
(the term pseudo for $C_{\mu}$ refers to its behaviour under
parity). This two potential approach has the advantage over
the Dirac string approach \cite{dirac} or the Wu-Yang fiber
bundle approach \cite{wu} in that it requires neither a singular
string variable nor a patching of the gauge potential. Using
two potentials also puts magnetic charge on the same footing
as electric charge by treating both as $U(1)$ gauge symmetries
\cite{carmeli}. The drawback of this approach is
that there are two ``photons'' in the theory rather than the
one photon that is observed \cite{hagen}. This can be overcome 
in two ways : Either by putting extra conditions on the two
gauge fields so that only the number of degrees of freedom 
necessary for one photon are left \cite{zwang}, or by accepting
the other ``photon'', but hiding it and the magnetic charge
associated with it through the Higgs mechanism \cite{singleton}.

The two potential theory of electric and magnetic charge allows
the dual symmetry of Maxwell's equations to be extended to the
level of the gauge fields. Using this with the 't Hooft-
Polyakov construction it is a trivial matter to construct
topologically stable electric poles rather than magnetic poles.
The major difference between the magnetically
charged soliton and an electrically charged soliton
is in the enormous difference of their respective
masses. We will first review the relevant aspects of the 
two potential theory.

\section{The Dual Four-Vector Potential}

In three-vector notation, Maxwell's equations with electric and 
magnetic charge are (in Lorentz-Heavyside units) \cite{jackson}
\begin{eqnarray}
\label{max1}
\nabla \cdot {\bf E} = \rho _e \; \; \; \; \nabla \times {\bf B} =
{1 \over c} \left( {\partial {\bf E} \over \partial t} +
{\bf J} _e \right ) \nonumber \\
\nabla \cdot {\bf B} = \rho _m \; \; \; \; - \nabla \times {\bf E} =
{1 \over c} \left( {\partial {\bf B} \over \partial t} +
{\bf J} _m \right) 
\end{eqnarray}
Introducing two four-vector potentials, $A ^{\mu} = (\phi _e ,
{\bf A})$ and $C^{\mu} = (\phi _m , {\bf C})$, the {\bf E} and
{\bf B} fields can be written as
\begin{eqnarray}
\label{ebtv}
{\bf E} = - \nabla \phi _e - {1 \over c} {\partial {\bf A} \over
\partial t} - \nabla \times {\bf C} \nonumber \\
{\bf B} = - \nabla \phi _m - {1 \over c} {\partial {\bf C} \over
\partial t} + \nabla \times {\bf A}
\end{eqnarray}
The usual definitions of {\bf E} and {\bf B} only involve $\phi _e$
and {\bf A}. Substituting the above expanded definitions for {\bf E}
and {\bf B} into Maxwell's equations, Eq. (\ref{max1}),  yields
(after using some standard vector identities and applying the Lorentz
gauge condition to both four-vector potentials) the wave equation
form of Maxwell's equations for both $A ^{\mu}$ and $C ^{\mu}$. The
equation for $A^ {\mu}$ has electric charges and currents ($J ^{\mu} _e
\equiv (\rho _e , {\bf J} _e)$) as sources, while the equation for
$C ^{\mu}$ has magnetic charges and currents ($J ^{\mu} _m \equiv 
(\rho _m , {\bf J} _m)$) as sources. In the two potential
theory all of Maxwell's equations are dynamical equations.

The two four-vector potentials, $A ^{\mu}$ and $C ^{\mu}$, are
similiar except for their behaviour under parity transformations.
The {\bf E} -field is an ordinary vector under
parity, and the {\bf B} -field is a pseudovector.
The normal definition of the fields in terms of the potentials,
implies that $\phi _e$ must a scalar and {\bf A} must be 
a vector under parity. In order for the {\bf E} and {\bf B} fields
to retain their parity properties under the expanded definitions
of Eq. (\ref{ebtv}), $\phi _m$ must be a pseudoscalar and {\bf C}
must be a pseudovector under parity. Therefore $A ^{\mu}$ is a
four-vector while $C ^{\mu}$ is a pseudo four-vector.

The two potential theory can be cast most simply in four-vector
notation. Defining the following two field strength tensors
\begin{eqnarray}
\label{fst}
F ^{\mu \nu} = \partial ^{\mu} A ^{\nu} - \partial ^{\nu} A ^{\mu}
\nonumber \\
G ^{\mu \nu} = \partial ^{\mu} C ^{\nu} - \partial ^{\nu} C ^{\mu}
\end{eqnarray}
and their duals
\begin{eqnarray}
{\cal F} ^{\mu \nu} = {1 \over 2} \epsilon ^{\mu \nu \alpha \beta}
F _{\alpha \beta} \nonumber \\
{\cal G} ^{\mu \nu} = {1 \over 2} \epsilon ^{\mu \nu \alpha \beta}
G _{\alpha \beta}
\end{eqnarray}
where $\epsilon ^{\mu \nu \alpha \beta}$ is the Levi-Civita
tensor, with $\epsilon ^{0123} = +1$ and having total antisymmetry
in its indices. The {\bf E} and {\bf B} fields then can be written as
\begin{eqnarray}
\label{eb1}
E _i = F^{i0} - {\cal G} ^{i0} = F ^{i0} + {1 \over 2} 
\epsilon ^{ijk} G_{jk} \nonumber \\
B _i = G ^{i0} + {\cal F} ^{i0} = G ^{i0} - {1 \over 2}
\epsilon ^{ijk} F_{jk}
\end{eqnarray}
Maxwell's equations in four-vector notation become
\begin{eqnarray}
\label{max2}
\partial _{\mu} F ^{\mu \nu} = \partial _{\mu} \partial ^{\mu}
A^ {\nu} = J ^{\nu} _e \nonumber \\
\partial _{\mu} G ^{\mu \nu} = \partial _{\mu} \partial ^{\mu}
C ^{\nu} = J ^{\nu} _m
\end{eqnarray}
where the Lorentz condition ($\partial _{\mu} A ^{\mu} =
\partial _{\mu} C ^{\mu} = 0$) has been imposed on the two
potentials in going from the first to the middle
expression. Finally the dual symmetry of Eqs. (\ref{dual1}),
(\ref{dual2}) can now be written in terms of the two four-vector
potentials and the two four-currents
\begin{eqnarray}
\label{dual3}
A ^{\mu} \rightarrow cos \theta A ^{\mu} + sin \theta C ^{\mu}
\; \; \; \; \;
C ^{\mu} \rightarrow -sin \theta A ^{\mu} + cos \theta C ^{\mu} 
\nonumber \\
J ^{\mu} _e \rightarrow cos \theta J^{\mu}_e 
+ sin \theta J ^{\mu} _m \; \; \; \; \;
J ^{\mu} _m \rightarrow -sin \theta J^{\mu}_e
+ cos \theta J^{\mu} _m
\end{eqnarray}
Eq. (\ref{dual3}) extends the dual symmetry of Eq. (\ref{dual1})
to the level of the four-vector potentials. This
implies that it should be possible 
to construct a topological electric charge in exactly the
same way 't Hooft and Polyakov constructed a topological
magnetic charge. The relevant parts of their solution are
reviewed in the next section.

\section{The 't Hooft-Polyakov Monopole Solution}

't Hooft and Polyakov \cite{thooft} independently discovered
the possibility of constructing a finite-energy, magnetically
charged soliton in a non-Abelian gauge theory coupled to a 
spontaneous symmetry breaking scalar field.
The stability of this field configuration was guaranteed by
the nontrivial homotopy of the Higgs field at spatial infinity
\cite{arafune}. 

In constructing the monopole solution 't Hooft considered an 
$SO(3)$ gauge theory coupled to a triplet scalar field with
the following Lagrange density
\begin{equation}
\label{lagrange1}
{\cal L} = - {1 \over 4} H_{\mu \nu} ^a H^{a \mu \nu} + {1 \over 2}
D_{\mu} \Phi ^a D ^{\mu} \Phi ^a + {1 \over 2} \mu ^2 \Phi ^a
\Phi ^a - {1 \over 4} \lambda (\Phi ^a \Phi ^a) ^2
\end{equation}
where
\begin{equation}
H _{\mu \nu} ^a = \partial _{\mu} W _{\nu} ^a - \partial _{\nu}
W_{\mu} ^a + g \epsilon ^{abc} W_{\mu} ^b W_{\nu} ^c
\end{equation}
and
\begin{equation}
D_{\mu} \Phi ^a = \partial _{\mu} \Phi ^a + g \epsilon ^{abc}
W_{\mu} ^b \Phi ^c
\end{equation}
$\epsilon^{abc}$ are the structure constants of $SO(3)$. The
$SO(3)$ gauge coupling, $g$, is at this point unspecified. If 
$\mu ^2 > 0$ and $\lambda > 0$ then the scalar field
develops a vacuum expectation value of $v = {\mu \over 
\sqrt{\lambda}}$. Inserting the following spherically 
symmetric ansatz into the equations of motion that come 
from the Lagrangian of Eq. (\ref{lagrange1})
\begin{eqnarray}
\label{ansatz}
W_i ^a &=& {\epsilon _{aij} x^j [1 - K(r)] \over g r^2} \; \;
\; \; \; W_0 ^a =0 \nonumber \\
\Phi ^a &=& {x^a H(r) \over g r^2}
\end{eqnarray}
one arrives at the coupled differential equations for
the functions $K(r)$ and $H(r)$
\begin{eqnarray}
\label{eqnmot}
r^2 K '' &=& K (K ^2 + H^2 -1) \nonumber \\
r^2 H '' &=& 2 H K^2 - \mu ^2 r^2 H \left( 1 - {\lambda \over
g^2 \mu ^2 r^2} H^2 \right)
\end{eqnarray}
where the prime means ${d \over dr}$. In addition to the
pure gauge solution to these equations (i.e. $K(r) =$ 0 and
$H(r) = {g \mu \over \sqrt{\lambda}} r$) there exists a nontrivial
finite energy solution. That such a solution exists can best be
seen by calculating the energy of the the field configuration
of Eq. (\ref{ansatz})
\begin{eqnarray}
\label{energy}
E &=& \int  T^{00} (r) d^3 x \nonumber \\
\bigskip
&=& \int \left({ 1 \over 4} H_{ij} ^a H^{aij} - {1 \over 2}
D_i \Phi ^a D ^i \Phi ^a  - {1 \over 2} \mu ^2 \Phi ^a
\Phi ^a + {1 \over 4} \lambda ( \Phi ^a \Phi ^a ) ^2 \right) 
d^3 x \nonumber \\
\bigskip
&=& {4 \pi \over g^2} \int _0 ^{\infty} \left( (K') ^2 +
{(K^2 -1)^2 \over 2 r^2} + {H^2 K^2 \over r^2} +
{(rH' - H)^2 \over 2 r^2} - {\mu ^2 H^2 \over 2}
+ {\lambda H^4 \over 4 g^2 r^2} \right) dr
\end{eqnarray}
Adding the constant term, $- {1 \over 4} \lambda v^4$, to the
Lagrangian, allows us to write the scalar field potential as
$ {1 \over 4} \lambda (\Phi ^a \Phi ^a - v^2) ^2$ which means
that the last two terms in Eq. (\ref{energy}) become
\begin{equation}
+ {\lambda \over 4} r^2 \left( {H^2 \over g^2 r^2} - v^2  \right) ^2
\end{equation}
Thus every term in Eq. (\ref{energy}) is positive-definite. Then, 
since neither $K(r) =$ 0 nor $K(r) =$ 1 is the lowest minimum, the 
variational principle requires that an intermediate
solution must exist. An analytical solution to Eq. (\ref{eqnmot})
has been found \cite{prasad} in the limit when both $\mu ^2$
and $\lambda$ are equal to zero. When $\mu ^2$ and $\lambda$
are non-zero the solution must be found numerically, and then
Eq. (\ref{energy}) becomes
\begin{equation}
\label{energy2}
E = { 4 \pi \over g^2} M_W f \left( {\lambda \over g^2}
\right)
\end{equation}
where $M_W = \mu g / \sqrt{\lambda}$ is the mass of two of
the $SO(3)$ gauge bosons after symmetry breaking. The function
$f({\lambda \over g^2})$, which must be evaluated numerically,
is ${\cal O} (1)$. 

To embed a $U(1)$ electromagentic symmetry into the $SO(3)$ theory,
the following gauge-invariant generalization of the 
Maxwell field strength tensor is defined
\begin{eqnarray}
\label{maxfst}
F_{\mu \nu} &=& \partial _{\mu} A _{\nu} - \partial _{\nu} A_{\mu}
- {1 \over g \vert \Phi \vert ^3} \epsilon _{abc} \Phi ^a
(\partial _{\mu} \Phi ^b) (\partial _{\nu} \Phi ^c) \nonumber \\
A _{\mu} &=& {1 \over \vert \Phi \vert} \Phi ^a W _{\mu} ^a
\end{eqnarray}
To see how a monopole emerges from this generalized field strength
tensor, the asymptotic values of the ansatz of Eq.
(\ref{ansatz}) are inserted into Eq. (\ref{maxfst}). As 
$r \rightarrow \infty$ $K(r) \rightarrow 0$ and $H(r) \rightarrow
{g \mu \over \sqrt{\lambda}} r$ which means
\begin{eqnarray}
\label{bcs}
W_i ^a \rightarrow \epsilon _{aij} x^j / g r^2
 + {\cal O} (r ^{-2}) \nonumber \\
\Phi ^a (r) \rightarrow  r^a v / r + {\cal O} (r ^{-2})
\end{eqnarray}
The asymptotic configuration of the scalar field is called 
the ``hedgehog'' solution, since the Higgs field approaches its
vacuum value $v$ in a peculiar way. Instead of pointing in
a fixed direction in isospin space for all points in 
confirguration space ({\it i.e.} $\rho ^a (r)= v \delta ^{a3} = 
v [0, 0, 1]$) it points in an isospin direction that coincides 
with the spatial radial direction. This links the internal (isospin)
space with the external (configuration) space. Inserting the
asymptotic fields of Eq. (\ref{bcs}) into the generalized 
field strength tensor of Eq. (\ref{maxfst}) yields
\begin{eqnarray}
\label{farfield1}
B_i &=& - {1 \over 2} \epsilon_{ijk} F^{jk} 
\rightarrow {r ^i \over g r^3} \nonumber \\
E_i &=& F ^{i0} = 0
\end{eqnarray}
So as $r \rightarrow \infty$ the fields rapidly approach those
of a Coulomb magnetic field and zero electric field. Taking
the non-Abelian gauge coupling to be equal to the usual ``electric''
$U(1)$ coupling ($g = e$) the magnetic charge implied by the
far fields of Eq. (\ref{farfield1}) is $4 \pi / e$.

\section {Electrically Charged Soliton}

The definition of the $U(1)$ field strength tensor 
and the identification of $W ^a _{\mu} \Phi ^a
/ \vert \Phi \vert$ with $A_{\mu}$, in Eq. (\ref{maxfst}),
are arbitrary. The term, $W ^a _{\mu} \Phi ^a / \vert \Phi \vert$,
could just as easily have been used to define the 
``magnetic'' gauge potential, $C _{\mu}$. The only change this
would require is that either the $SO(3)$ gauge fields, 
$W ^a _{\mu}$, or the scalar fields, $\Phi ^a$, would have to
be pseudo quantities, since $C _{\mu}$ is
a pseudo four-vector. Using the dual symmetry of
Eq. (\ref{dual3}), the two four-vector potentials 
can be transformed into one another by taking $\theta =
{\pi \over 2}$. This gives $A_{\mu} \rightarrow C_{\mu}$, and
from Eq. (\ref{fst}) it also changes the ``electric'' field
strength tensor into the ``magnetic'' field strength tensor
($F _{\mu \nu} \rightarrow G _{\mu \nu}$). In this way
Eq. (\ref{maxfst}) becomes
\begin{eqnarray}
\label{maxfst2}
G _{\mu \nu} &=& \partial _{\mu} C _{\nu} - \partial _{\nu} C _{\mu}
- {1 \over g \vert \Phi \vert ^3} \epsilon _{abc} \Phi ^a 
(\partial _{\mu} \Phi ^b) (\partial _{\nu} \Phi ^c) \nonumber \\
C _{\mu} &=& {1 \over \vert \Phi \vert} \Phi ^a W _{\mu} ^a
\end{eqnarray}
As with the expression for the ``electric'' field strength tensor,
$G_{\mu \nu}$ is gauge-invariant. Inserting the
asymptotic values of the non-Abelian gauge fields and of the scalar
fields (which now define the magnetic gauge potential, $C_{\mu}$)
from Eq. (\ref{bcs}) into Eq. (\ref{maxfst2}) one finds that
the far fields of this soliton are
\begin{eqnarray}
\label{farfield2}
B_i &=& G ^{i0} = 0
\nonumber \\
E_i &=& {1 \over 2} \epsilon ^{ijk} G _{jk} \rightarrow
- {r ^i \over g r^3}
\end{eqnarray}
where the expanded definitions of the {\bf E} and {\bf B}
fields from Eq. (\ref{eb1}) have been used. Requiring that the
charge of the {\bf E} field in Eq. (\ref{farfield2}) have the
magnitude of the charge of an electron (or proton) leads to
the requirement that $g = 4 \pi / e$ (where $e$ is the magnitude
of the electron's charge). This means that the original 
$SO(3)$ coupling, $g$, must be large. In the case of the 
magnetic soliton the $SO(3)$
gauge coupling was taken to be $g = e$, since there one wanted 
to embed the electric $U(1)$ symmetry (with gauge boson, $A_{\mu}$,
and coupling, $e$) into the non-Abelian theory. In the present case,
the coupling of the magnetic $U(1)$ symmetry, which is embedded in
the $SO(3)$ theory, is fixed by the condition that the  
electric charge of the soliton be that of observed particles.

The main difference between the electric soliton and the 
magnetic soliton is in their masses. By equating the energy
in the fields with the mass of the soliton one finds, from
Eq. (\ref{energy2}), that the magnetically charged soliton
(which has $g=e$) has a mass of
\begin{equation}
\label{massmag}
M_m = E = {4 \pi \over e^2} M_W f \left( {\lambda \over e^2}
\right) \approx 137 M_W
\end{equation}
since numerically $f({\lambda \over e^2})$ is ${\cal O} (1)$.
In contrast the electric soltion (which has $g = 4 \pi / e$) has
a mass of 
\begin{equation}
\label{massele}
M_e = E = {e^2 \over 4 \pi} M_W f \left( \lambda e^2  \right)
\approx {1 \over 137} M_W
\end{equation}
The function, $f (\lambda e^2)$, is ${\cal O} (1)$ since
the energy of the fields (Eq. (\ref{energy})) is invariant under
the duality transformation that turns the the magnetic soliton
into an electric soliton. One might worry that
the argument of $f$ is different in the two cases. It can be
shown that $f(0) =$ 1 \cite{prasad}, and increases monotonically
with the argument. Thus for a given $\lambda$  the
argument of the electric case is always closer to zero and the 
value of the function $f$ is closer to 1. From Eqs. (\ref{massmag}), 
(\ref{massele}) it is seen that the mass of the electric soliton 
is over $10 ^4$ times smaller than that of the magnetic soliton.
If the mass of the gauge boson, $M_W$, is taken to be of the order
of the electroweak  gauge bosons ({\it i.e.} ${\cal O} (100)$ GeV)
then such an electrically charged, spin zero particle should have
already been detected. This would seem to imply that if such
electric solitons exist, the non-Abelian gauge group, into
which they are embedded, must undergo symmetry breaking in such
a way that the mass of the gauge bosons are at least
several orders of magnitude greater then the masses of the
electroweak gauge bosons. Even if $M_W$ were in the range of
50 TeV it might be possible to see such an electric soliton
at some reasonably extrapolated future accelerator. An
alternative possibility would be to use the spin from isospin 
mechanism \cite{rebbi} and form bound states out of particles 
with various combinations of topological electric charge
and gauge magnetic charge. These bound states would carry
a spin of 1/2, obey Fermi-Dirac statistics \cite{goldhaber},
and be in the mass range of the baryons. In this paper, however,
we simply want to show the theoretical possibility of obtaining
a topological electric charge from a non-Abelian gauge theory,
since the $SO(3)$ group which is used here is apparently not a
theory picked by nature. We will leave for a future work the
task of building a more realistic model through the use of a
larger non-Abelian symmetry.

In standard electrodynamics only 
the {\bf B} field can be written as a curl. Including magnetic
charge in electrodynamics as a gauge charge, 
by introducing a second four-vector potential,
then requires that part of the {\bf E} field be given by
the curl of this second potential. 
The crucial element in constructing a finite-energy, stable
field configuration, with either a Coulomb electric or
magnetic far field, is being able to write that field as
the curl of some vector potential. A general argument
can be given \cite{li} that shows this. In order for the
energy of the soliton, Eq. (\ref{energy}), to be finite the 
covariant derivative of the scalar field must satisfy 
the following boundary condition as $r \rightarrow \infty$
\begin{eqnarray}
\label{bcs2}
D_{\mu} \Phi ^a &=& \partial _{\mu} \Phi ^a + g
\epsilon ^{abc} W_{\mu} ^b \Phi ^c \nonumber \\
&\rightarrow & {\cal O} (r ^{-2})
\end{eqnarray}
In order for there to be a Coulomb far field (either electric
or magnetic) the gauge fields, $W_{\mu} ^a$, must go like
$r ^{-1}$ as $r \rightarrow \infty$. In addition
the magnitude of the scalar field must approach a constant 
(its VEV) as $r \rightarrow \infty$. Then even though each
of the two seperate terms need not approach zero like 
$r ^{-2}$, some cancellation can occur between the
terms such that $D _{\mu} \Phi ^a \rightarrow 0$ 
like $r ^{-2}$ . This is what happens with
the ansatz of Eq. (\ref{ansatz}). For time-independent
fields the time component of the first
term of Eq. (\ref{bcs2}) is zero, so no cancellation can occur
between the two terms. Therefore $ W ^a _0$ must go to zero
faster than $r ^{-1}$ and does not give rise to a
Coulomb far field. When the $U(1)$ gauge field is identified
with $W_{\mu} ^a \Phi ^a / \vert \Phi \vert$, as in Eqs. 
(\ref{maxfst}) or (\ref{maxfst2}), this implies that the
time component of the $U(1)$ gauge field also will not
yield a Coulomb far field. In standard electrodynamics, where
the {\bf E} field is defined only by $F ^{i0}$, a Coulomb
field is only possible if $A_0 \ne$ 0 (in fact if $A_0 =$ 0
and only static solutions are considered then ${\bf E}
=$ 0). In the two potential theory, however, the {\bf E} field
also has a part that is the curl of a vector potential
({\it i.e.} $E_i = 1/2 \epsilon ^{ijk} G_{jk}$). A
Coulomb far field is then possible if the spatial components
of the non-Abelian gauge field (and therefore the spatial
components of the embedded $U(1)$ gauge field) go to zero like
$r ^{-1}$, as is the case for the ansatz of Eq. (\ref{ansatz}). 

Looking in detail at where the Coulomb far fields come
from, it appears as if they are due entirely to
the scalar fields. Inserting the asymptotic field conditions of
Eq. (\ref{bcs}), into the generalized field strength tensors 
of Eqs. (\ref{maxfst}) or (\ref{maxfst2}), it is found that the 
Coulomb fields come only from the last term of the generalized
field strength tensors, which involve only the scalar fields.
This makes it appear that whether the soliton has a magnetic 
charge or an electric charge is completely independent
of the type of $U(1)$ gauge field (either $A_{\mu}$ or $C_{\mu}$) 
that is embedded into the non-Abelian gauge theory. This is not 
the case. It has been shown \cite{arafune} that by performing
a gauge transformation  to the unitary or Abelian gauge, it
is possible to rotate the scalar field into the more common
asymptotic vacuum configuration
\begin{equation}
\Phi ^a (r) \rightarrow v \delta ^{a3} = v (0, 0, 1)
\end{equation}
Using this asymptotic scalar field in the generalized 
field strength tensors gives
\begin{eqnarray}
\label{agm}
G_{\mu \nu} &=& \partial _{\mu} W_{\nu} ^3 - \partial _{\nu}
W_{\mu} ^3 \nonumber \\
C_{\mu} &=& W _{\mu} ^3
\end{eqnarray}
or
\begin{eqnarray}
\label{age}
F_{\mu \nu} &=& \partial _{\mu} W_{\nu} ^3 - \partial _{\nu}
W_{\mu} ^3 \nonumber \\
A_{\mu} &=& W _{\mu} ^3
\end{eqnarray}
where the $U(1)$ gauge bosons are now associated exclusively with
the third isospin component $SO(3)$ gauge boson. Since the expression
for the generalized field strength tensors is gauge-invariant, one
still gets a Coulomb far field. This comes about, even though
the field strength tensors of Eqs. (\ref{agm}) (\ref{age}) 
are of the form that usually preclude the existence of a 
Coulomb field coming from the spatial part of the tensor, 
because the gauge transformation that takes the ``hedgehog''
gauge to the Abelian gauge is singular along the positive 
z-axis. In this way a connection between the `t Hooft-Polyakov
monopole and Dirac's monopole can be seen.

\section{Conclusions}

In this paper it has been shown that it is possible to construct
a finite-energy, topologically stable soliton with an electric
charge. This is accomplished through the application of the
't Hooft-Polyakov monopole solution to the two potential theory
of electric and magnetic charge, where the two types of charges
are treated as gauge charges through the introduction of two
four-vector potentials. In the case of the magnetically charged
soliton one starts with some non-Abelian gauge theory which is
coupled to a scalar field that breaks the gauge symmetry. By
embedding the electric $U(1)$ symmetry in the non-Abelian theory,
through the introduction of a generalized field strength tensor,
and taking the scalar field to go to the ``hedgehog'' solution
as $r \rightarrow \infty$, it is found that a stable, finite-energy,
magnetically charged soliton emerges. The gauge coupling, $g$, of
the non-Abelian group is required to satisfy, $g = e$, in order that
the embedded $U(1)$ symmetry may be identified with the usual 
electric $U(1)$ gauge field. This leads to both the magnetic charge
of the monopole ($ = 4 \pi / e$) and the mass of the monopole ($\sim
4 \pi / e^2$) being extremely large. The magnetic charge carried
by the soliton is not a Noether charge but a topological charge,
which is due to the nontrivial topology of the scalar field, rather
than from a symmetry of the Lagrangian. Thus by starting with an
electric gauge charge and using the 't Hooft-Polyakov ansatz
one ends up with a topological magnetic charge.

Based on the dual symmetry between electric and magentic quantities
it is reasonable to believe that this construction can be reversed
-- {\it i.e.} start with a magnetic gauge charge and using
the `t Hooft-Polaykov ansatz, end up with a topological electric
charge. This is in fact trivially possible if magnetic charge is
treated, like electric charge, as a gauge symmetry by introducing
the pseudo four-vector potential, $C_{\mu}$. Using the dual
transformation in terms of the potentials to
``rotate'' the electric potential, $A_{\mu}$, into the magnetic
potential, $C_{\mu}$, it is found that the magnetic soliton gets
transformed into an electric soliton. Requiring that the electric
charge of this soliton be equal in magnitude to the charge
of other electrically charged particles ({\it e.g.} electrons,
protons) we found that the original non-Abelian gauge coupling
must satisfy $g = 4 \pi / e$. This made the mass of the electrically
charged soliton several orders of magnitude lighter than its 
magnetic counterpart ($M_e \sim {1 \over 137} M_W$ compared to 
$M_m \sim 137 M_W$). Taking $M_W$ to be of the order of the
electroweak gauge boson masses, such an electrically charged,
spin zero particle should have been observed. This might be taken
to imply that such electric solitons are only of theoretical
interest. However by using the spin from isospin mechanism
\cite{rebbi} it may be possible to construct bound states of
topological electric charge and gauge magnetic charge, which
behave like spin 1/2 fermions \cite{goldhaber}, and have masses
roughly in the range of the baryon masses. Here, however, our goal 
was simply to show that it is possible to get a topological 
electric charge from a non-Abelian gauge theory, since the $SO(3)$
group is currently not thought to play a fundamental role in
particle physics.

Julia and Zee \cite{zee} have expanded the ansatz of
Eq. (\ref{ansatz}) by setting the time component of the $SO(3)$
gauge field equal to $ x^a J(r) / g r^2 $. In this way they found
that the field equations yielded a solution having both magnetic
and electric charge ({\it i.e.} a dyon). However the Julia and Zee
solution does not allow for the existence of an electric charge
without the presence of a magnetic charge. In addition the 
topological stability arguments \cite{arafune} that apply to 
the pure magnetic or electric solution do not apply to the
dyonic solution, although plausiblity arguments for its stability
are given.

\section{Acknowledgements} The author would like to thank
Hannelore Roscher and Justin O' Neill for useful suggestions
and encouragement during this work.

\end{document}